# On the telescopic disks of stars – a review and analysis of stellar observations from the early 17th through the middle 19th centuries

Christopher M. Graney

*Jefferson Community & Technical College, 1000 Community College Drive, Louisville, KY 40272 (USA)*

502-213-7292

christopher.graney@kctcs.edu

Timothy P. Grayson

*Independent Scholar*

timotpg@gmail.com

Since the dawn of telescopic astronomy astronomers have observed and measured the "spurious" telescopic disks of stars, generally reporting that brighter stars have larger disks than fainter stars. Early observers such as Galileo Galilei interpreted these disks as being the physical bodies of stars; later observers such as William Herschel understood them to be spurious; some, such as Christian Huygens, argued that stars show no disks at all. In the early 19th century George B. Airy produced a theoretical explanation of star images sufficient to explain all historical observations, but astronomers were slow to fully recognize this. Even today conventional wisdom concerning stars and telescopes stands at odds to both historical observations and Airy's theory.

We give a detailed analysis of both historical observations and Airy's theory, illustrating how Airy's theory explains the historical observations, from Galileo to Huygens to Herschel. We argue that the observations themselves appear in all cases to be valid and worth further study.

222

# Introduction

The "spurious" telescopic disks of stars have been observed and measured since the very beginning of telescopic astronomy. And since that beginning a consistent theme has been present in the work of those astronomers who possessed both the observing skills and the quality instruments needed to see those disks – brighter stars have larger disks than fainter stars. Early observers such as Galileo (Finocchiaro 1989: 167-168, 173-174, 180) interpreted these disks as being the physical bodies of stars. Later observers like Halley (1720: 3) and Herschel (1805: 40-44) understood the telescopic disks of stars to be semi-spurious – they were partly a product of the telescope, but partly a product of the star itself. When George Airy produced a theoretical explanation for these disks (1835: 283-291), astronomers in the mid-19th century were slow to reconcile this theory, which said that all stellar images were characterized only by the aperture of the telescope in question, with their long-standing observations that the diameters of the disks varied with magnitude, even though Airy's discussion fully explained that variation. Even today, the conventional wisdom about stars seen in telescopes stands at odds to both historical observations and Airy's theory. However, an in-depth analysis of diffraction of light reveals that diffraction theory not only accounts for the general theme that brighter stars have larger disks than fainter stars, but also accounts for some of the finer points of detail in observations of telescopic disks of stars by visual astronomers. This in turn leads the authors to conclude that the historical observations of telescopic disks of stars discussed in this paper – including pioneering observations by 17th century astronomers Simon Marius, Galileo Galilei, Christian Huygens, and Johannes Hevelius – are valid and reliable data. This is despite that a casual review of these astronomers' work seems to suggest that their reports on stars are in disagreement. But it is the astronomers' interpretations of the observations that can be erroneous, due in large part to their ignorance of diffraction theory. The observations themselves should not be considered rhetorical enhancements, "thought observations" made for the sake of argument, or products of wishful thinking and crude instruments. Historical observations of stars may even be of value to modern astronomers.

# The physical bodies of stars: Simon Marius and Galileo Galilei

March of 2010 marked the 400th anniversary of Galileo Galilei's publication of his *Sidereus Nuncius* – his first astronomical publication. The *Nuncius* includes a brief discussion of Galileo's observations of the stars in which he writes about a difference between the appearances of the planets and that of the fixed stars:

> …the fixed stars do not look to the naked eye bounded by a circular circumference, but rather like blazes of light, shooting out beams on all sides and very sparkling, and with a telescope they appear of the same shape as when they are viewed by simply looking at them [Galilei, Kepler, Carlos 1880: 40]....

Four years later, Simon Marius, in his *Mundus Iovialis*, reported that stars did show telescopic disks (Dreyer 1909: 191). In the *Mundus* Marius challenges Galileo's statements in the *Nuncius* –

> …not only the planets but also all the more conspicuous fixed stars are discerned to be clearly round, and especially the bright stars of Canis Major, Minor, Orion, Leo, Ursa Major, etc.... I am truly surprised that Galileo with his quite excellent instrument has never seen this. For instance, he writes in his *Sidereus Nuncius*, that the stars do not possess a marked circular border [Marius 1614: 46-48]....[1]

Marius's note that telescopic disks of stars are most obvious in the brightest stars marks the first appearance of the "brighter stars have larger disks than fainter stars" theme that will recur in stellar observations for the next two and a half centuries.

Galileo soon joined Marius in observing telescopic disks of stars. In 1617 he observed the star Mizar in the tail of Ursa Major (Fedele 1949; Ondra 2004; Siebert 2005). He found Mizar to be double, and he recorded differing diameters for its two components – 6 seconds and 4 seconds of arc (Favaro 1890: III, 877). In 1617 he also made a precise sketch (Figure 1) of the Trapezium in Orion (Graney 2008: 262), and noted that the closely-spaced stars in the Trapezium had varying diameters (Favaro 1890: III, 880).  In 1624, in a lengthy letter to Francesco Ingoli, he discussed stars at length, saying that they were round and typically measured 4 seconds in diameter –

> …if you measure Jupiter's diameter exactly, it barely comes to 40 seconds, so that the sun's diameter becomes 50 times greater; but Jupiter's diameter is no less than ten times larger than that of an average fixed star (as a good telescope will show us), so that the sun's diameter is five hundred times that of an average fixed star [Finocchiaro 1989: 167]....

---

[1] Translation from Latin to English by C. Graney.

He goes on to discuss how stars could range from somewhat larger for bright stars to considerably smaller for faint stars visible to the naked eye, with stars too faint to see with the naked eye and visible only with the telescope being smaller still (Finocchiaro 1989: 167-168, 174, 180).  In his 1632 *Dialogue Concerning the Two Chief World Systems*, he again mentions stars as having measurable disks, with brighter stars having larger disks and fainter stars having smaller disks:

> …the apparent diameter of a fixed star of the first magnitude is no more than 5 seconds, or 300 thirds, and the diameter of one of the sixth magnitude measures 50 thirds [359]....

He even proposes to make measurements that involve obscuring half of a star's disk, something he feels is achievable with the proper step-up and a good telescope (388).

Both Marius and Galileo interpreted the telescopic disks of stars as being the stars' physical bodies.  Each believed that the disks yielded information about stellar distances.  Interestingly, Marius concluded that the disks were evidence in favor of a Tychonic world system (Marius 1614: 48; Graney 2010), while Galileo used them to argue for a Copernican world system (Galilei 1632: 358-360; Finocchiaro 1989: 166-180).

Other astronomers attempted to measure the telescopic disks of stars.  Johannes Hevelius published a table giving the diameters of 19 stars in his *Mercurius in Sole Visus Gedani* of 1662; he listed the largest star in the table, Sirius, as measuring 6 seconds and 21 thirds of arc, and the smallest, a sixth-magnitude star in Orion, as measuring 1 second and 56 thirds (94).  In 1717 Jaques Cassini determined Sirius to have a diameter of 5 seconds (545).

But not all astronomers believed stars to have measurable disks.  Christiaan Huygens argued that stars were merely points of light, as evidenced by their decreasing in size when a smoked glass was placed in the light path of a telescope (1659: 7; Roberts 1694).[2]

# "Optick Fallacies" and spurious disks:  Halley and William Herschel

By the early 18th century the spurious nature of the telescopic disks had become apparent to some.  Edmund Halley wrote a commentary in 1720 on Cassini's observation of Sirius.  Cassini had stopped down the aperture of his telescope to an inch and a half in order to make his measurement of the star's disk, which he interpreted as being Sirius's physical body.  About this measurement Halley wrote –

> …it may not perhaps be amiss to enquire whether the suppos'd visible Diameter of Sirius were not an Optick Fallacy, occasioned by the great contraction of the Aperture of the Object Glass:  For we all know that the Diameters of Aldebaran and Spica Virginis, are so small, that when they happen to immerge on the dark Limb of the Moon, they are so far from loosing their Light gradually, as they must do were they of any sensible magnitude, that they vanish at once with their utmost Lustre; and emerge likewise in a Moment, not small at first, but at once appear with their full Light, even tho' the Emersion happen very near the Cusps; where if they were four Seconds in Diameter they would be many Seconds of Time in getting entirely separated from the Limb.  But the contrary appears to all those, that have observed the occultations of these bright Stars.  And tho

---

[2] Huygens' argument was not necessarily convincing to early astronomers who observed the telescopic disks of stars.  For example, Flamsteed argued that the disks were real, and that the soot on Huygens' glass merely allowed thin rays of light to pass through, rendering the appearance of smaller size (Bailey 1835: 206-207).

> Sirius be bigger than either of them, yet he is by far less than two of them; and consequently his Diameter to theirs is less than the Square Root of 2 to 1, or than 14 to 10; whence in Mr. Cassinni's excellent 36 Foot Glass, those Stars ought to be about four Seconds in Diameter; and they would undoubtedly appear so if viewed after the same manner [1720: 3]....

Note that while Halley views telescopic disks of stars to be spurious, he still states that a brighter star (Sirius) will have a larger disk than fainter ones (Aldebaran, Spica), at least when "viewed after the same manner".

William Herschel shared Halley's views. His 1782 catalog of double stars includes information about "the comparative size of the stars" (112). For example, his catalog description of γ Andromeda (130) reads

> \* γ Andromedae, FL. 57. Supra pedem sinestrum. August 25, 1779. Double. Very unequal. L.[larger] reddish w.[white]; S.[smaller] fine light sky-blue, inclining to green… A most beautiful object.

γ Andromeda's components in fact differ by almost three magnitudes, with the brighter being of spectral class K and the fainter being of class A. They are reasonably described as reddish white and light sky-blue. Herschel likewise describes β Cygni as a "considerably unequal" double with the larger star being pale red and the smaller star being "a beautiful blue" (142-143). The components of β Cygni are in fact less unequal in magnitude than γ Andromeda, with the brighter being pale red and the fainter being blue. His description of the Trapezium in Orion (129-130) makes an interesting comparison to Galileo's sketch (Figure 1):

> θ Orionis, FL. 41. Trium contiguarum in longo ensis media. November 11, 1776. Quadruple. It is the small telescopic Trapezium in the Nebula; Considerably unequal. The most southern star of the following side of the Trapezium is the largest; the star in the opposite corner [which Galileo did not see] is the smallest; the remaining two are nearly equal…. The two stars in the preceding side distance 8".780; in the southern side 12".812; in the following side 15".208; in the northern side, 20".396.

Herschel understood that star sizes depend on the specifics of the observation, even though they show some consistency in that the brighter (larger in his words) star in a pair is always larger, regardless of conditions.  In 1805 Herschel published a list of conclusions concerning stellar disks seen through telescopes. He states that that the disks of stars are spurious, but under the "same circumstances" their dimensions are permanent.  He recognizes that altering the optical system, such as changing the aperture, or significantly changing magnification, affects the disks.  While the changes are not simple relationships, in general:  increasing aperture decreases disk size; obscuring the light path can reduce disk size dramatically (Herschel discusses observing the disk of Arcturus reducing in size as the skies became progressively more hazy); increasing magnification increases the apparent disk size, but the change is less than what would be predicted by the increase in magnification, as seen in Figure 2 (Herschel 1805: 40-42).   But he concludes that

> …many causes will have an influence on the apparent diameter of the spurious disks of the stars; but they are so far within the reach of our knowledge, that with a proper regard to them, the conclusion we have drawn… that under the same circumstances 'their dimensions are permanent,' will still remain good [44].

## Reconciling telescopic disks with theory: John Herschel and Airy

The 19th century saw the introduction of a theory that could explain the telescopic disks of stars – the undulatory (wave) theory of light.  A generation after William Herschel described the various characteristics of the disks, his son John Herschel contributed a lengthy article on light to the *Encyclopedia Metropolitana* (1828). In the article John Herschel discusses in detail the appearance of a bright star as seen through a telescope, and notes what he sees as a shortcoming of the undulatory theory – namely that it fails to account for the size variation with brightness of the telescopic disks of stars.  He begins by describing stars seen at low magnification, and his description sounds much like Galileo's from the *Sidereus Nuncius*:

> When we look at a bright star through a very good telescope with a low magnifying power, its appearance is that of a condensed, brilliant mass of light, of which it is impossible to discern the shape for the brightness; and which, let the goodness of the telescope be what it will, is seldom free from some small ragged appendages or rays [491].

But then he describes the image seen at higher magnification:

> But when we apply a magnifying power from 200 to 300 or 400, the star is then seen (in favourable circumstances of tranquil atmosphere, uniform temperature, &c.) as a perfectly round, well-defined planetary disc, surrounded by two, three, or more alternately dark and bright rings, which, if examined attentively, are seen to be slightly coloured at their borders. They succeed each other nearly at equal intervals round the central disc [491]....

He goes on to state that –

> These discs were first noticed by Sir William Herschel, who first applied sufficiently high magnifying powers to telescopes to render them visible [491].

– which, as we have seen, is not true. Next he discusses the nature of the disks, with some reference to the undulatory theory of light:

> They are not the real bodies of the stars, which are infinitely too remote to be ever visible with any magnifiers we can apply; but spurious, or unreal images, resulting from optical causes, which are still to a certain degree obscure. It is evident, indeed, to any one who has entered into what we have said of the law of interferences, and from the explanation given in Art. 590 and 591 of the formation of foci on the undulatory system, that (supposing the mirror or object-glass rigorously aplanatic) the focal point in the axis will be agitated with the united undulations, in complete accordance, from every part of the surface, and must, of course, appear intensely luminous; but that as we recede from the focus in any direction in a plane at right angles to the axis, this accordance will no longer take place, but the rays from one side of the object-glass will begin to interfere with and destroy those from the other, so that at a certain distance the opposition will be total, and a dark ring will arise, which, for the same reason, will be succeeded by a bright one, and so on. Thus the origin both of the central disc and the rings is obvious, though to calculate their magnitude from the data may be difficult [491].

John Herschel then brings up the matter of the apparent size of the disks varying with magnitude, which he feels the undulatory theory fails to explain:

> But this gives no account of one of the most remarkable peculiarities in this phenomenon, viz. that the apparent size of the disc is different for different stars, being uniformly larger the brighter the star. This cannot be a mere illusion of judgment; because when two unequally bright stars are seen at once, as in the case of a close double star, so as to be directly compared, the inequality of their spurious diameters is striking; nor can it be owing to any real difference in the stars, as the intervention of a cloud, which reduces their brightness, reduces also their apparent discs till they become mere points. Nor can it be attributed to irradiation, or propagation of the impression from the point on the retina to a distance, as in that case the light of the central disc would encroach on the rings, and obliterate them; unless, indeed, we suppose the vibrations of the retina to be performed according to the same laws as those of the ether, and to be capable of interfering with them; in which case, the disc and rings seen on the retina will be a resultant system, originating from the interference of both species of undulations [491].

Lastly, he enters into a discussion of the appearance of stars when the aperture of the telescope is reduced:

> When the whole aperture of a telescope is limited by a circular diaphragm, whether applied near to, or at a distance from, the mirror or object-glass, the disc and rings enlarge in the inverse proportion of the diameter of the aperture. When the aperture was much reduced (as to one inch, for a telescope of 7 feet focal length) the spurious disc was enlarged to a planetary appearance, being well denned, and surrounded by one ring only, strong enough to be clearly perceived, and faintly tinged with colour in the following order, reckoning from the centre of the disc. White, very faint red, black, very faint blue, white, extremely faint red, black. When the aperture was reduced still farther (as to half an inch) the rings were too faint to be seen, and the disc was enlarged to a great, size, the graduation of light from its centre to the circumference being now very visible, giving it a hazy and cometic appearance, as in [Figure 3; 492].

A few years later, in 1835, George Biddell Airy wrote a thorough discussion of the undulatory theory applied to image formation by a telescope – his "On the Diffraction of an Object-glass with Circular Aperture". Airy's discussion answers John Herschel's article – Airy specifically mentions it (290) – and provides a succinct explanation for all observations of telescopic disks of stars discussed here.

After working through the requisite mathematical treatment of light diffracting through a telescope's circular aperture he discusses what the mathematics tells us. Within this discussion Airy uses *a* for the radius of the aperture of the telescope in inches, and "0.000022 inches" as the mean wavelength of light:

> The image of a star will not be a point but a bright circle surrounded by a series of bright rings. The angular diameters of these will depend on nothing but the aperture of the telescope, and will be inversely as the aperture [287].

He goes on to say that "the central spot has lost half its light" when the distance *s* measured from the center of the "bright circle" is *s=1.17/a*, and that "there is total privation of light, or a black ring", when *s=2.76/a*. "The brightest part of the first bright ring corresponds to" *s=3.70/a*, where "its intensity is about 1/57 of that at the center". "[T]here is a black ring when" *s=5.16/a*, and "the brightest part of the second bright ring corresponds to" *s=6.09/a*, with intensity of 1/240 the center. He says the intensity of the third bright ring is about 1/620 of the center (287-288). See Figure 4.

He then goes on to answer John Herschel:

> The rapid decrease of light in the successive rings will sufficiently explain the visibility of two or three rings with a very bright star and the non-visibility of rings with a faint star. The difference of the diameters of the central spots (or spurious disks) of different stars (which has presented a difficulty to writers on Optics) is also fully explained. Thus the radius of the spurious disk of a faint star, where light of less than half the intensity of the central light makes no impression on the eye, is [*s=1.17/a*], whereas the radius of the spurious disk of a bright star, where light of 1/10 the intensity of the central light is sensible, is determined by *s=1.97/a* [Figure 5].

The general agreement of these results with observation is very satisfactory [288].

This pattern of illumination produced by light from a point source diffracting through a circular aperture that Airy is describing is known today as an "Airy pattern". The agreement of these results with the observations discussed in this paper is indeed satisfactory.

Airy's results applied to Galileo's observations of Mizar have yielded good agreement (Graney 2007). They have yielded truly remarkable agreement when applied to the Hevelius table of star diameters (Graney 2009). In Figure 6 we apply Airy's results to Cassini and Halley's work and find that Halley's statement, that a telescope of 1.5 inch aperture that produces a 5 second image of Sirius should produce a 4 second image of Aldebaran or Spica, is quite reasonable. Airy's results agree with William Herschel's description of disks being affected by changes in aperture, magnification, etc. (Figure 7). And lastly, they even agree with William and John Herschel's discussions of star disks vanishing to points when haze or a cloud intervenes, and with Huygens' assertion that stars do not have disks at all based on observations through a smoked glass (Figure 8).

## Confusion regarding Airy's work: Spurious disks and tiny points

Many astronomers did not grasp the comprehensive nature of Airy's work in "On the Diffraction of an Object-glass with Circular Aperture". How and even whether the undulatory theory accounted for the variation with brightness of the sizes of telescopic disks of stars was not considered a settled matter.

Thus we find in 1867 G. Knott reporting in the *Monthly Notices of the Royal Astronomical Society* that he used a spherical crystal micrometer and a 7 1/3 inch Alvin Clark refractor stopped down to various apertures to measure the telescopic disks of a number of stars so as to test Airy's theory. He produced a table of star diameters not unlike that of Hevelius two centuries earlier, but for varying apertures. Like Hevelius's data, Knott's data shows star diameters decreasing with magnitude. It also shows star diameters increasing with decreasing aperture (88). Then a decade later we find a letter to *The Observatory* by George Hunt quoting John Herschel's encyclopedia article on light and concluding that the undulatory theory did not account not for observations because the diameter of the telescopic disk was a function not only of aperture –

> but also, in some *unexplained* manner, of the relative light of those stars, conventionally termed their magnitude [emphasis added; Hunt 1879: 152].

To this the editors of *The Observatory* responded –

> We think that the variation in the size of the spurious disk according to the brightness of the star may be explained by the circumstance that, according to the Undulatory Theory, the light fades away *gradually* from the central point outwards to the first dark ring, and that with the fainter stars it is only the central portion which is sufficiently bright to produce a sensible impression. Sir G. Airy has not given the diameter of the spurious disk, but that of the first dark ring, which is its extreme limit [emphasis original; Hunt 1879: 152-153].

We see the issue of the sizes of telescopic disks of stars and the undulatory theory continue to be a point of debate into the later decades of the 19th century, to become intertwined with debate about the sizes of photographically recorded disks, as can be seen in a debate concerning the sizes of telescopic disks of stars seen visually and recorded photographically that is recorded in a yet later volume of *The Observatory* ("Meeting of the Royal Astronomical Society" 1886).

And despite the comprehensive nature of both John Herschel's and Airy's respective discussions concerning the appearance of stars as seen through a telescope, a great deal of erroneous or at least greatly oversimplified information about the disks has circulated since in the wider world of popular astronomy. The statement that stars seen through a telescope are mere points which no amount of magnification can affect has appeared in otherwise reliable sources over a great span of time. The 1875 *Handbook of Astronomy* states that stars seen through a telescope are mere lucid points having no detectable size regardless of the level of magnification used. The *Handbook* adds that William Herschel used magnifications as high as six thousand, and that according to Herschel, such magnifications actually made the sizes of stars seem even less, "if possible" (379; how could anything be smaller than a point?)[3] Lardner was an astronomy professor from the University of London; Dunkin served on the staff of the Royal Observatory and would be elected president of the Royal Astronomical Society. The same sentiment can be found today:

> Even though most stars in the nighttime sky are bigger and brighter than the Sun, nothing but a tiny point is seen telescopically; no matter how large the instrument or what magnification is used to observe a star [Dickenson 1988: 83-84].

Or:

> Why are stars so special? Simply because no amount of magnification will make them appear bigger – all their light remains tightly concentrated into a tiny point [Seronik 2010].[4]

Or:

> Stars will appear like twinkling points of light in the telescope. Even the largest telescopes cannot magnify stars to appear as anything more than points of light! [Orion ShortTube].[5]

In fairness to these sources, recall J. Herschel's statement that –

> When we look at a bright star through a very good telescope with a low magnifying power, its appearance is that of a condensed, brilliant mass of light, of which it is impossible to discern the shape for the brightness; and which, let the goodness of the telescope be what it will, is seldom free from some small ragged appendages or rays [1828: 491].

Telescopic disks are only visible in larger aperture telescopes at high magnifications, or (keeping in mind Airy's work showing that disk size varies inversely with aperture), in small aperture telescopes at more modest magnifications. It has been a long time since serious observers used small aperture telescopes, and few astronomers look at stars with very high magnifications the way William Herschel did. The conventional wisdom is that "telescopes do not magnify stars," and that is valid enough for many people. The details discussed in J. Herschel and Airy's work are not commonly understood.

---

[3] Apparently the *Handbook* is erroneously interpreting Herschel's ideas about the effect of magnification on stars (see Figure 2) as meaning that their disks actually reduce in size as it appears to the eye.

[4] Seronik is an associate editor of *Sky & Telescope* magazine and writes its "Binocular Highlight" and "Telescope Workshop" features.

[5] Orion is a major seller of amateur telescopes.

## Why we should care about telescopic disks of stars

If conventional wisdom states that telescopes do not magnify stars then it is not surprising to find the concept appearing in the work of historians. Stillman Drake, in his translation of the *Sidereus Nuncius*, footnotes Galileo's description of fixed stars appearing as "blazes" with the comment:

> Fixed stars are so distant that their light reaches the earth as from dimensionless points. Hence their images are not enlarged by even the best telescopes, which serve only to gather more of their light and in that way increase their visibility [1990: 47 note 16].

Misconceptions about the telescopic appearance of stars may strike the reader as not being an issue of consequence. However, such misconceptions hinder progress in the history of astronomy and even in astronomy itself, leading to dead-ends, hiding potentially productive areas of investigation, and hiding valuable historical data.

Absent misconceptions about the telescopic appearance of stars, could Paul Feyerabend have formulated his criticism of Galileo (and by extension science in general)? The words of Feyerabend, "science's worst enemy" (Preston, Munévar, Lamb D. 2000), can still stir up controversy, as seen in the 2008 squabble at La Sapienza University in Rome (BBC News 2008). But Feyerabend's criticism is grounded in the conventional wisdom that telescopes do not magnify stars, and so is a dead-end. So will be any analysis that leans heavily on the conventional wisdom and does not recognize the telescopic disks of stars.[6]

Would the disks of stars seen through a telescope have altered the calculus that says that the telescope supported the heliocentric Copernican world system, by eliminating Tycho Brahe's objection that stars sufficiently distant to satisfy Copernicus would be ridiculously large? Simon Marius thought so. He argued that the telescopic disks of stars that he saw endorsed the geocentric Tychonic world system (Marius 1614: 48; Graney 2010). Yet if historians think of stars in terms of the conventional wisdom, who would think to investigate such a line of thought as the telescope supporting Tychonic geocentrism?

Could the disks of stars reported by observers such as Hevelius constitute photometric data of some precision that might be of value to modern astronomers? Historical data such as naked-eye estimates of magnitude have been used by astronomers before (Shara, Mofatt, Webbink 1985). Disk data from Hevelius, Herschel, and perhaps others could be of great interest to astronomers (Graney 2009). However, since the story told by historians, following the conventional wisdom about stars seen through telescopes, is that no precise method of photometry existed prior to W. Herschel,[7] astronomers would be unlikely to look for such data.

---

[6] Feyerabend is selected here simply as a recognizable example. A discussion of the presence of the conventional wisdom in the writings of historians and philosophers of science would occupy a full paper. However, a brief list of sources will help to illustrate the point:
Feyerabend includes the subject of magnification of stars, couched in terms of the conventional wisdom, in his *Against Method* (92-93), including a lengthy footnote that refers to Herschel talking about stars not responding to dramatic increases in magnification – see Figure 2. In critiquing Feyerabend, Alan Chalmers (1985) brings up the issue of Galileo's telescopes not magnifying stars (176). Feyerabend aside, we find other writers working based on the conventional wisdom. Harold Brown (1985) discusses the issue of stars not being magnified (490). Henry Frankel's 1978 paper on Galileo's non-telescopic star observations includes quotes from Galileo making reference to disks of stars (77-80), but also includes quotes from other writers (78 note 6), including Dreyer (1906: 414) and Kuhn (1985: 221), who state the conventional wisdom. An understanding that stars show disks in small telescopes, which are magnifiable with defined sizes of a few seconds, and which were seen by Marius and Galileo, might affect these various writers' analyses.

And lastly, there is intangible value in simply having the story right. The conventional wisdom does not reflect what historical astronomers were seeing. It would be best if the conventional wisdom were put to rest.

## Technical Questions

At this point some technical questions may be arising in the reader's mind that we can anticipate and address. One such question is the issue of such small disk sizes being reported by observers using very small aperture telescopes. Do they not violate known optical principles concerning the resolving power of even "optically perfect" ("diffraction-limited") telescopes, and are therefore suspect?

For example, Cassini used a telescope stopped to 1.5 inch aperture in order to measure the telescopic disk of Sirius to be 5 seconds of arc in diameter, and Halley states that Aldebaran or Spica should have disks of 4 seconds diameter seen with the same telescope – a difference of a mere second. However, the

---

[7] Historians write that the telescope provided no help in measuring the magnitudes of stars:
> Part of the problem lay in the lack of a sufficiently delicate technique for monitoring the apparent brightness of a star. Stars were simply grouped according to the crude classification inherited from Antiquity, whereby the brightest stars were first magnitude and the faintest, sixth. The mid-nineteenth century would see the invention of new instruments to give an objective measure of the brightness of stars, and a new definition of magnitude. But before then, in the closing years of the eighteenth century, astronomers were at last provided with a simple method of determining whether a star had in fact altered in brightness [this being William Herschel's "Catalogues of the Comparative Brightnesses of Stars"; Hoskin 1997: 201-202]

Hearnshaw (1996: 12-13) and Miles (2007: 173) attribute to the telescope no opportunity for improvement in photometry. Yet as Airy illustrated, disk diameter reflects the brightness of stars. It can, and at least in the case of Hevelius, did serve as a means of distinguishing magnitude to some precision (Graney 2009). In similar manner to what is mentioned in the previous footnote on Feyerabend, etc., an understanding that stars show disks in small telescopes, might affect these various writers' analyses.

Rayleigh criterion, which states that the resolving power of a telescope is defined by Airy's first dark ring, should be no better than 3.7 seconds,[8] even if the telescope is optically perfect. Likewise, Galileo claims to be able to see star disk sizes as small as 2 seconds (Finocchiaro 1989: 174), even though his telescopes probably had an aperture of roughly an inch and therefore a resolving power of 5.5 seconds even if optically perfect.

The answer to this question is that the resolving power of a telescope relates to its ability to distinguish real detail in real objects. The telescopic disks of stars, being spurious, do not involve resolution of real detail – they are artifacts of the instrument. The Cassini/Halley disk data only states that a difference in diameter of 1 second can be distinguished in these spurious disks, which does not violate optical principles. Had Cassini claimed to be able to distinguish two stars separated by 1 second, then that would indeed violate the Rayleigh criterion. Likewise, had Galileo claimed to be able to see stars separated by 2 seconds that would violate the Rayleigh criterion, but his claim to be able to distinguish a 2 second disk does not.

A second technical question involves how astronomers – especially early astronomers using "Galilean" style telescopes that form no real image, have no focal plane, and therefore cannot have a reticle or similar measuring device – could measure such small disks. Should Marius and Galileo truly be included with later astronomers as having seen the telescopic disks of stars and observed and measured brighter stars to have larger disks? Marius only provides a qualitative description of telescopic disks, not actual measurements, so all that is required is for him to be able to discern them. Galileo does report explicit measurements for Mizar, and general measurements for stars as a whole. Drake and Kowal (1980) have discussed Galileo's technique for making measurements with the telescope. They have discussed that Galileo detected Neptune with a telescope, and noticed its motion over a one-day period; that Galileo correctly recorded a change in the diameter of Jupiter of just over two seconds of arc, from 41.5 seconds to 39.25 seconds. Standish and Nobili (1997) have determined that Galileo recorded the positions of Jupiter's moons to precisions of 0.1 Jovian radii (approximately 2 seconds of arc), and Graney (2007) has shown his sketch of the Trapezium to be to perhaps even better accuracy.  Therefore, Galileo's measurements of Mizar's disks as being 6 seconds and 4 seconds are not inconsistent with his overall work (see Figure 9).  Less analysis has been done of Marius's work, but he measured the diameter of the Andromeda nebula, discerned it as having a dull, pale light which increased in brightness toward its center (Bond 1848: 75-76), and described it as being like "a candle shining through horn" (Watson 2005: 86).   He also found that he was only able to reconcile his observations of Jupiter's moons to his calculations if he factored in Jupiter's motion about the Sun (in a Tychonic fashion; Prickard 1916: 404, 408-409).   And he observed the location of Tycho Brahe's supernova of 1572 and found a star there which he estimated to be "somewhat dimmer than Jupiter's third moon" (Waldrop 1988: 462).  That he noticed problems with Jupiter's moons, and then was able to reconcile them with calculations, suggests that he too must have developed a system of precise measurement, perhaps similar to Galileo's.  His precise comparison of a star to a moon of Jupiter suggests a system for photometry – perhaps disk size measurement?  At any rate, in both the case of Marius and the case of Galileo, broad evidence for skill at measurement with the telescope exists.

---

[8] using Airy's $s=2.76/a$ formula.

A final question is, as Airy's work assumes diffraction-limited optics, is it not absurd to apply Airy's ideas to 17th century telescopes that were certainly not "optically perfect"?  No, it is not.  First, 17th century optics were not necessarily poor – the optics of telescopes attributed to Galileo have been tested and found to indeed be "optically perfect" (Greco, Molesini,  Quercioli 1992).  Secondly, deviations from optical perfection do not automatically mean Airy's work does not apply.  A functional but not optically perfect telescope will still produce the basic image structure described by Airy's paper – the location of the first dark ring, for instance, will not be greatly changed, nor will the ring not be dark.  The central disk remains distinct, and of the same form.  There is some loss of intensity in the central disk, but deviations from optical perfection that don't render the telescope unusable will not change what has been discussed here (Mahajan 2001: 129-130, 135).

## Conclusions

We have reviewed visual observations of the telescopic disks of stars from Simon Marius in 1614 through the rise of stellar photography in the late 19th century. We have shown that there is a broad consistency in these observations – consistency between early observers and late observers, and consistency between all observers and the theory of George Biddell Airy. A hallmark of these observations, including even the earliest, is the theme that brighter stars have larger disks than fainter stars. But even the observations of an astronomer who argued against the existence of stellar disks, Christian Huygens, are consistent with Airy's theory.

This broad consistency suggests that the data on stars recorded by these observers should be considered reliable. This includes the data gathered by the earliest observers, Marius and Galileo. Even though they may have erroneously attributed their observations to measurements of the physical bodies of stars, their empirical data were nevertheless reliable and accurate. This reliable data in turn implies that these observers had the requisite quality of instruments and observing skill required to observe such elusive phenomena as the telescopic disks of stars. Ideas about telescopes rendering stars as unmagnifiable, and historically uninteresting, points of light need to be set aside. There may be much of interest in the telescopic disks of stars – both for historians of astronomy and for astronomers themselves.

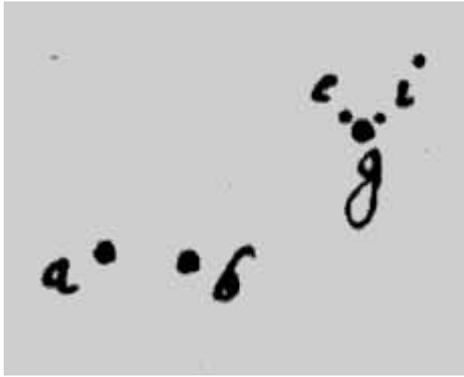

Fig. 1

Galileo's sketch of the Trapezium (trio in upper corner) and two other stars (Favaro 1890: III, 880).

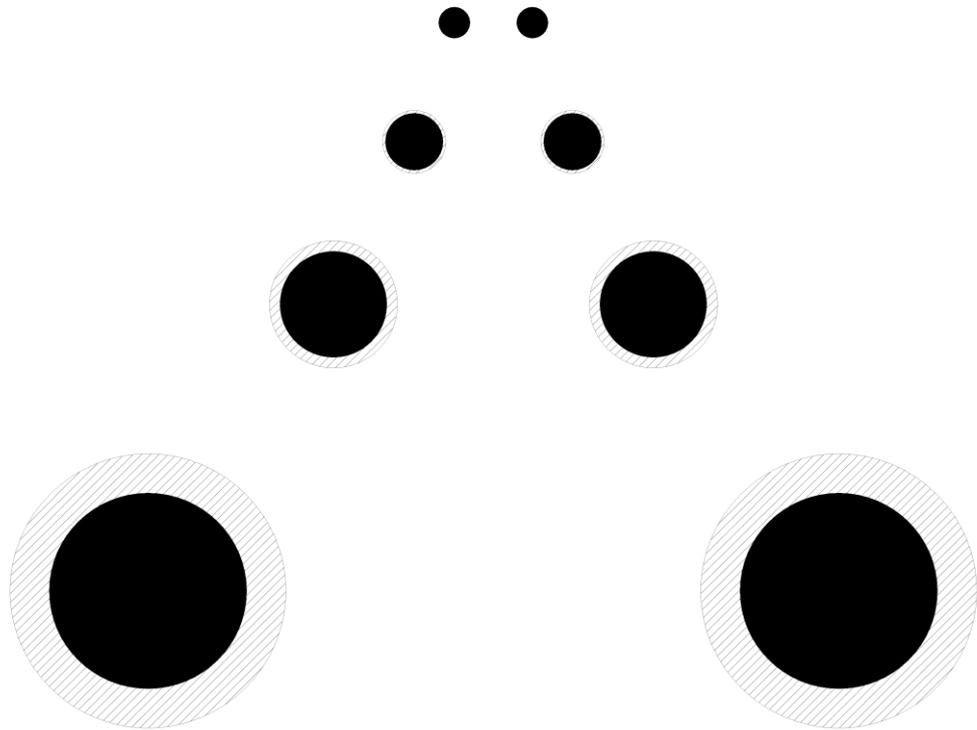

Fig. 2

Diagram showing how a double star with equal components (part of ε Lyrae) appeared to Herschel under increasing magnification. He reports that the interval between the stars [shown top to bottom in diagram] "with a power of 227 is almost 1½ diameter of either; with 460, full 1¾ diameter; with 932, 2 diameters ; with 2010, 2½ diameters. These estimations are a mean of two years observations [1782: 123]." The dark disks in the diagram are the diameters Herschel reported. The hatched disks are the diameters the stars would have if they scaled proportionately with magnification. The first doubling of magnification produces a more or less proportionate increase in disk size, but more dramatic increases in magnification do not enlarge the disks as much as would be expected. As the magnification of a telescope is increased, the apparent size of one arc second of angle increases proportionately. Since the center-to-center separation of stars is a fixed number of arc seconds, their separation in the telescope image increases proportionately to magnification. But since the appearance of the spurious disks does not grow as fast as the magnification, they are reducing in size as measured by arc seconds. This is strong indication that the disks cannot be tied to a real physical object, since a real physical object cannot shrink by increasing telescope magnification. This is something Herschel notes he sees in many observations of double stars (Herschel 1805, pg. 41-42). Note that the disks do enlarge with magnification, and for modest changes the "reducing" effect would not be apparent.

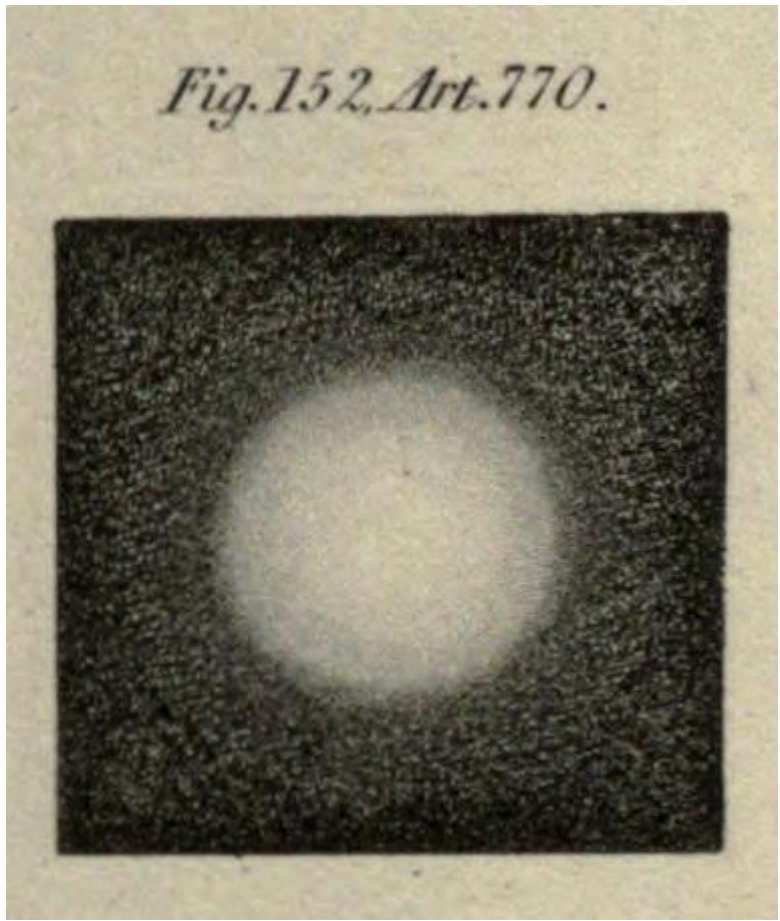

Fig. 3

John Herschel's illustration of a star as seen through a small aperture telescope, showing a clearly defined spurious disk (from the *Encyclopedia Metropolitana*).

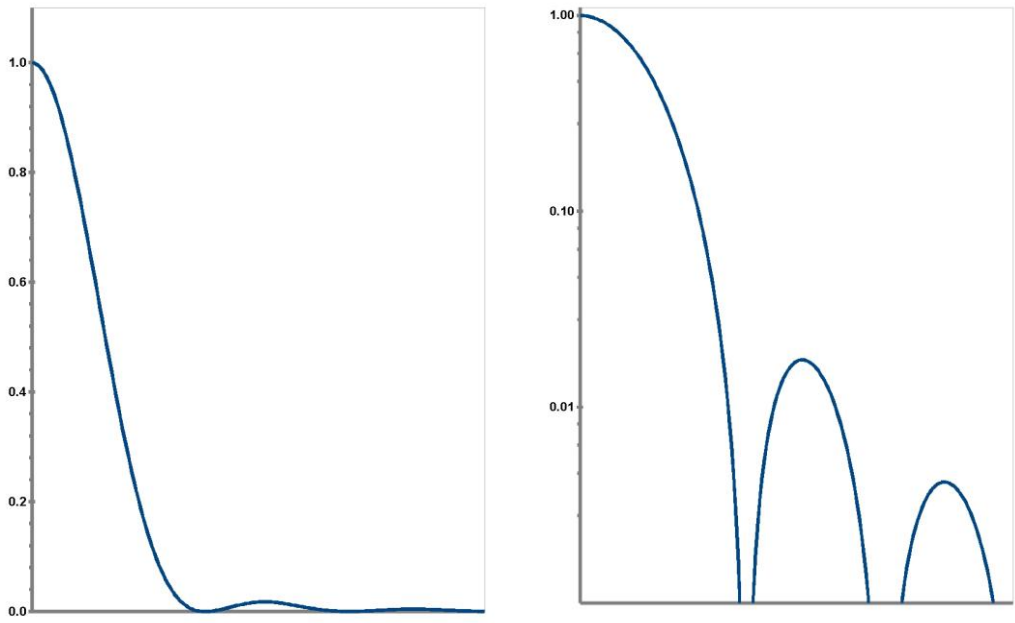

Fig. 4

Linear axis (left) and log axis (right) plots of intensity in the circular aperture diffraction pattern that is the image of a star formed by a telescope.

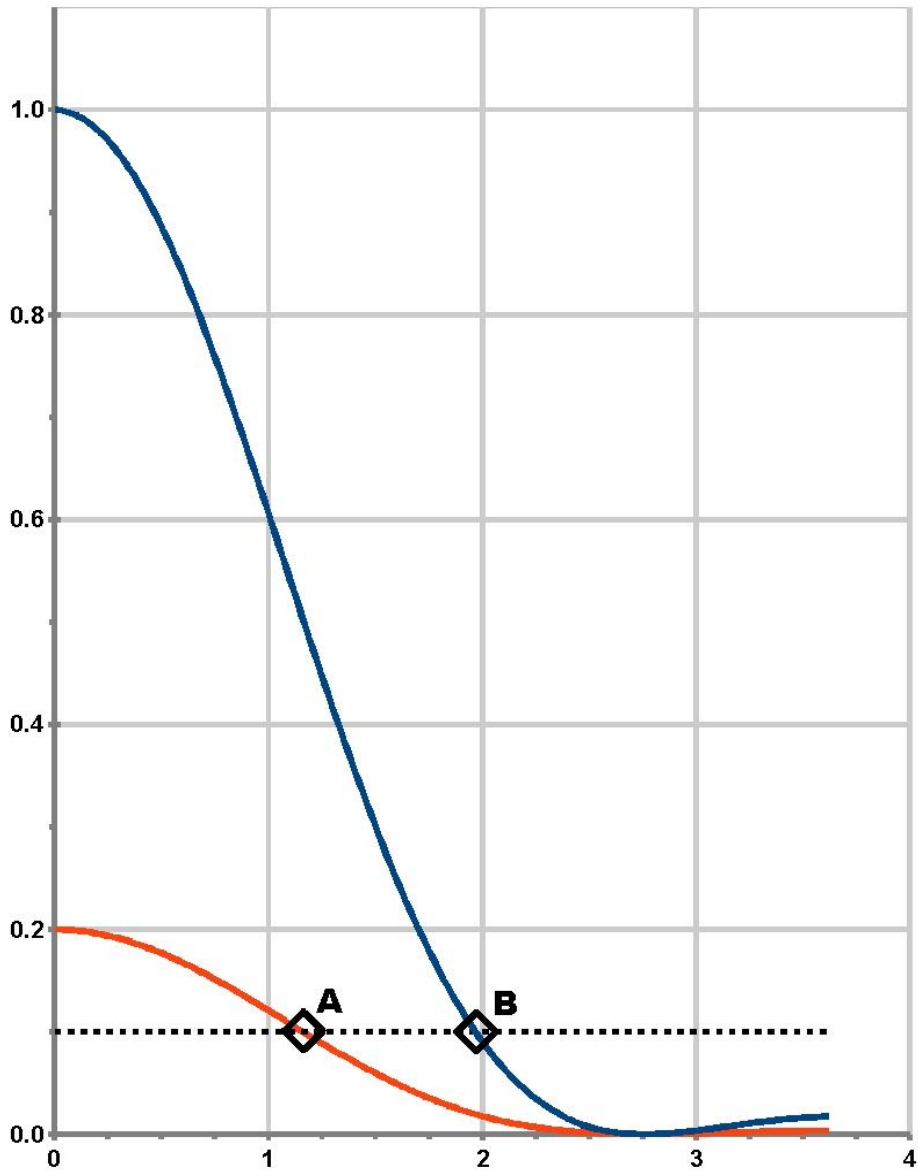

Fig. 5

Airy's discussion of the radius of a star "where light of less than half the intensity of the central light makes no impression on the eye" (lower curve), and a star where "light of 1/10 the intensity of the central light is sensible". A and B mark the radii, which Airy gives as 1.17 seconds and 1.97 seconds for a telescope with aperture of 1 inch radius. The horizontal dashed line represents Airy's threshold of sensitivity – light more intense than the threshold is 'sensible', whereas light less intense is not. This plot is on a linear axis – the horizontal axis being seconds of arc and the vertical being relative intensity.

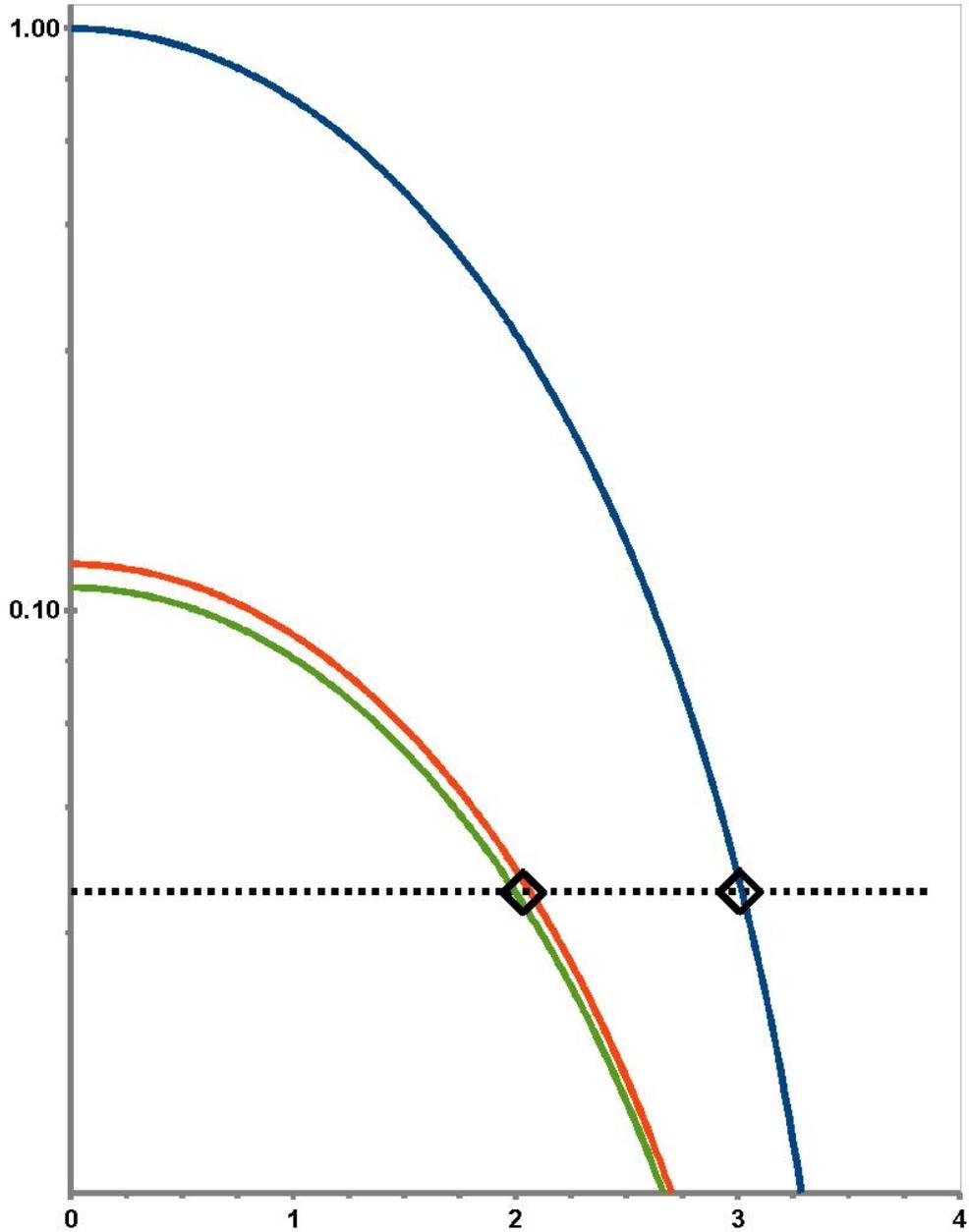

Fig. 6

Log axis plots of intensity curves for Aldebaran and Spica (lower curves) and Sirius for a telescope with aperture 1.5 inch in diameter. The horizontal axis is seconds of arc and the vertical axis is relative intensity. A threshold of sensitivity selected such as to render the first two stars to have diameters of 4 seconds yields a 6 second diameter for Spica. Halley had estimated the values would be 4 and 5 seconds.

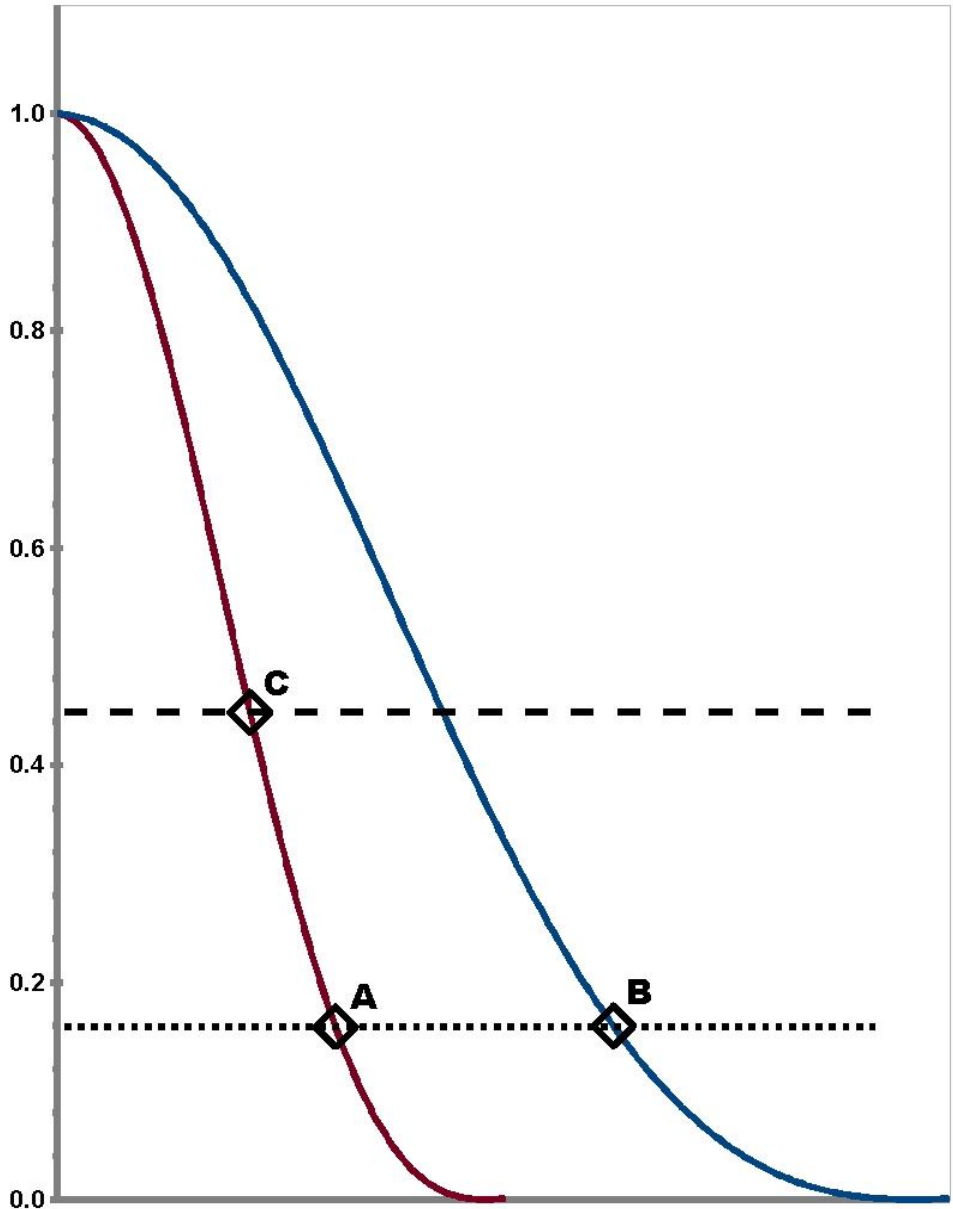

Fig. 7

Illustration of how changes in the optical system will cause the changes in the sizes of spurious disks of stars seen by William Herschel. A star seen through a telescope's full aperture has a certain radius (A), but if that aperture is reduced the diffraction pattern broadens, and the star radius increases (B). Likewise, if increasing the magnification raises the threshold of sensibility (for example, because the light is now spread over a wider area of the retina of the eye) the result will be a smaller star radius (C). This diagram is only intended to serve as a basic illustration. Changing the optical system by changing aperture or magnification will result in many changes. For example, reducing the aperture not only broadens the diffraction pattern, it reduces the light entering the telescope, and therefore the overall intensity of the pattern (which is not shown here). Changing magnification, such as by changing eyepieces, can result in a better or worse quality lens entering the system. Furthermore the threshold will vary significantly based upon many factors, such as the vision of the astronomer, his level of fatigue, and the amount of ambient light. The eye has two types of light-sensitive cells, one far more sensitive than the other, and so a single threshold as envisioned by Airy and shown in Figure 5 is probably an oversimplification. In short, direct observations of disk diameter made by eye will exhibit dependence upon many factors, and should vary from astronomer to astronomer and even from day to day for a given astronomer, although as can be seen in William Herschel's work, such variations are probably small. Nonetheless, William Herschel makes the point that such measurements must be made "under the same circumstances".

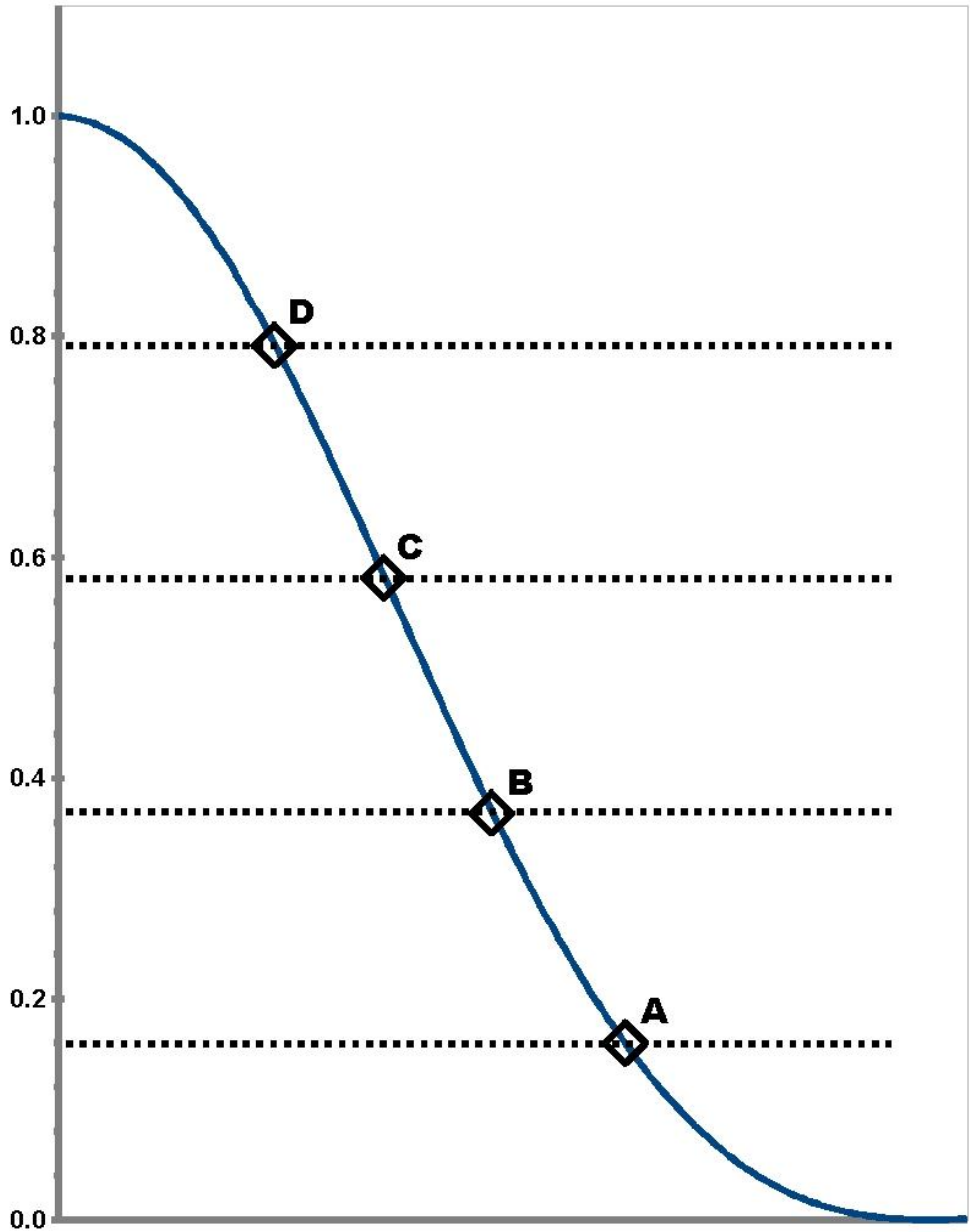

Fig. 8

If the threshold of sensitivity is raised, then the radius of the spurious disk of a star is reduced – from A, to B, to C, to D – eventually to nothing. Thus Huygens' method of reducing the light of a star with a smoked glass yields the result that stars have no measurable disks. William and John Herschel's reports of the spurious disks of stars shrinking as haze or clouds obscured the star are examples of the same phenomenon.

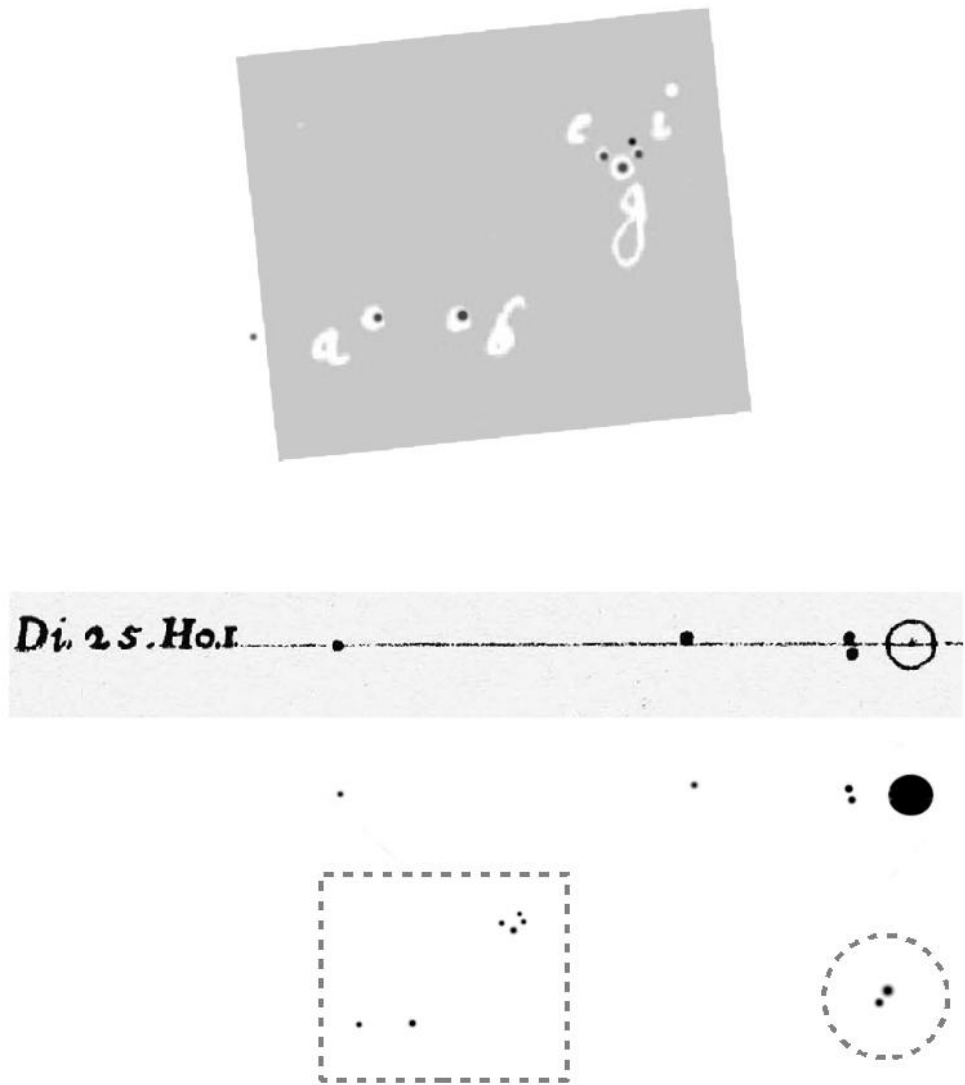

Fig. 9

Illustrations of the accuracy of Galileo's work. Top image is Galileo's sketch of the Trapezium from Figure 1 (negative – light gray region) superimposed on a modern chart of the same stars. The next image is one of Galileo's sketches of Jupiter from March 25, 1613 (Favaro 1890: V, 243), compared with a diagram of the positions of Jupiter's moons for that date and time created using the software package *Stellarium* (approximate time was taken from information provided by Galileo; exact time was determined by matching the positions of the moons). Last are representations of the Trapezium (boxed) and Mizar (circled) shown at the same scale as the Jupiter diagram. The star sizes in the Mizar representation are approximately correct to the sizes Galileo gives in his notes on Mizar. Comparing Mizar to the Jupiter diagram illustrates the relative difficulty of recording the separation and sizes of the components of Mizar as compared to recording the sizes and positions in the Jovian system. The star sizes in the Trapezium representation and the dots that represent Jupiter's moons in the Jupiter diagram may not be properly scaled and should not be taken as representing the sizes of those bodies as seen by Galileo.